# A comparative study of *ab initio* nonradiative recombination rate calculations under different formalisms

Lin Shi[1], Ke Xu[1], and Lin-Wang Wang[2]*

[1]Suzhou Institute of Nano-Tech and Nano-Bionics, Chinese Academy of Sciences, Suzhou 215125, People's Republic of China

[2]Materials Sciences Division, Lawrence Berkeley National Laboratory, One Cyclotron Road, Mail Stop 66, Berkeley, California 94720, USA

*lwwang@lbl.gov

## Abstract:

Nonradiative carrier recombination is of both great applied and fundamental importance. But the correct *ab initio* approaches to calculate it remains to be inconclusive. Here we used 5 different formalisms to calculate the nonradiative carrier recombinations of two complex defect structures GaP:$Zn_{Ga}$-$O_P$ and GaN:$Zn_{Ga}$-$V_N$, and compared the results with experiments. In order to apply different multiphonon assisted electron transition formalisms, we have calculated the electron-phonon coupling constants by *ab initio* density functional theory for all phonon modes. Compared with different methods, the capture coefficients calculated by the static coupling theory are $4.30 \times 10^{-8}$ and $1.46 \times 10^{-7}$ cm$^3$/s for GaP:$Zn_{Ga}$-$O_P$ and GaN:$Zn_{Ga}$-$V_N$, which are in good agreement with the experiment results, $\left(4_{-1}^{+2}\right) \times 10^{-8}$ and $3.0 \times 10^{-7}$ cm$^3$/s respectively. We also provided arguments for why the static coupling theory should be used to calculate the nonradiative decays of semiconductors.

# Introduction:

Nonradiative carrier recombination as often described by the Shockley-Read-Hall (SRH) [1-3] phenomenological model is a very important process in semiconductor physics. Normally nonradiative transitions reduce device efficiencies by reducing photo generated carriers, suppressing luminescence, reducing carrier lifetimes, or enhancing defect diffusion during device operations [4]. Since the direct measurement of such processes is often difficult [5-7], especially to identify the responsible defects, it is very desirable to use *ab initio* calculations to study the related phenomena. However, so far, there is a lack of commonly accepted way to calculate the SRH, which is the focus of the current study: to compare different approaches to establish the correct *ab initio* procedure.

The SRH recombination includes two carrier capture processes: first the photo generated, or the injected, minority carrier is captured by the recombination center, and then the majority carrier recombines with this defect center. The first step is often the rate determining step. For deep centers, the nonradiative carrier capture occurs via multiphonon emission (MPE) [8]. Many researchers have contributed to the theoretical foundations of MPE over the past six decades. These studies have revealed that the results of calculations are very sensitive to the adopted theoretical models, with different approaches yielding capture rates many orders of magnitudes different [9]. So far, there is a lack of systematic studies to compare different methods, and to find which method is most suitable for semiconductor SRH calculations. Part of the reason is due to the lack of efficient *ab initio* approach to calculate all the electron-phonon coupling constants often required in these formalisms. Recently we have proposed a variational approach to calculate all the electron-phonon coupling constants in a single self-consistent field (SCF) calculation [10]. In the current paper, we take the advantage of that method to comparatively study different SRH calculation formalisms. Some of these formalisms have been used by other researchers, and the others are proposed by us as plausible methods for SRH calculations.

The different formalisms studied in the current paper include i) Static coupling theory [11-13], ii)Adiabatic approximation [14], iii) Marcus theory [15], iv) Quantumcharge transfer (CT) theory [16] and v) one-dimensional (1D) formulism [17].

Recently, we have used the adiabatic approximation formalism (method ii) [10] to calculate the nonradiative recombination rate for GaN:$Zn_{Ga}$-$V_N$. Despite the fact that the *ab initio* calculated electron-phonon coupling constants and all the phonon modes are used in the formalism, the calculated rate is several orders of magnitude too small, especially compared to the newly measured experimental data [18] which appeared after Ref.[10]. Alkauskas et al. [17] developed another practical approach to calculate the nonradiative carrier capture coefficients (method v). They considered only one special collective phonon mode along the transition degree of freedom, and used it to replace the sum over all vibrational modes. In this 1D model, the atomic degree of freedom is represented by a single generalized configuration coordination $Q$. They also calculated the carrier capture rate of GaN:$Zn_{Ga}$-$V_N$, and found their result (at room-temperature) as $1.0 \times 10^{-8}$ cm$^3$/s, which is higher than our previous result $B_p = 5.57 \times 10^{-10}$ cm$^3$/s[10], but is still lower than the experiment results $B_p = 3 \times 10^{-7}$ to $3 \times 10^{-6}$ cm$^3$/s [18]. In their discussion, they argued for the use of static coupling theory instead of adiabatic formalism. Indeed, there was a long lasting debate, especially in 70s and early 80s about different methods to calculate nonradiative recombination processes [9,14,19]. It was found that the static coupling formalism yields results order of magnitudes larger than the adiabatic approximation formalism, and is generally in better agreements with experiments. In section II, we will come back to this point, and present our view regarding this issue.

The nonradiative recombination process is also a charge transfer (CT) process, which can be calculated by the classical Marcus theory [15] (method iii). Marcus theory has been widely used to study the charge transfer from one localized state to another localized state at different locations. For example, McKenna and Blumberger [20] have determined the electron transfer rate between two localized defect states in MgO using Marcus theory, while Tarafder et al [21] calculated the rate of hole transfer from a photo excited CdSe/CdS core/shell nanorod to a tethered ferrocene molecule with Marcus theory. However, most of these cases involve two localized states at two different locations, which make it possible to define the so called diabatic states (states localized at different places) [20,21]. It is more challenging to use the Marcus theory to calculate the nonradiative recombination process since the electronic states before and after the transition are localized at the same place (while one is extended, another is localized). Nevertheless, Henry and Lang in their seminal paper [8] has used essentially the Marcus theory to study the nonradiative recombination rate, and seemed to get good results when compared with experiments. But their calculations were not *ab initio*, instead effective mass models, and a few adjustable parameters are used to get the correct result. Here we will use *ab initio* methods to carry out the Marcus theory calculation for nonradiative recombination rate. The Marcus theory is a classical 1D theory, since it describes the atomic degree of freedom by a single reaction/transition coordinate Q. In the conventional Marcus theory for charge transfer between two localized states, the electron-electron coupling is not induced by phonon modes, instead it exists in the original zero phonon Hamiltonian between the two localized diabatic states (which are not electron eigen states to begin with, hence the coupling exists). The Marcus theory can also be extended using quantum mechanical treatments for the phonon degree of freedom while keeping the original electron-electron coupling constant. We call such formalism, as derived by Nan et al. [16] as quantum CT method (method iv). Compared to the classical Marcus theory, the quantum mechanical theory can have different temperature dependence, especially at low temperatures; it can provide the quantum tunneling effects for the phonon modes yielding larger transition rates than the classical formula.

To judge the accuracies of the different calculation results, we need to choose nonradiative recombination centers with unambiguous experiment results. The electron capture cross section for Zn-O center in GaP has been measured unambiguously to be $\left(2^{+2}_{-1}\right) \times 10^{-15}$ cm$^2$ at 300K by Jayson et al. in their excellent study [22]. The same system has also been studied by Henry and Lang, which proved that the carrier capture process in this system is indeed due to multiphonon emission [8]. We will also calculate GaN:$Zn_{Ga}$-$V_N$ complex. Not only it has been studied before in our previous work [10], there are also recent new experiments [18] which make the experimental results more certain.

# Static coupling and adiabatic approximation formalisms

Under the adiabatic approximation, the initial and final electron and phonon wave functions are described by Born-Oppenheimer approximations:

$$\Psi_{i,n}(r,R) = \psi_i(r,R)\varphi_{i,n}(R)$$
$$\Psi_{j,m}(r,R) = \psi_j(r,R)\varphi_{j,m}(R) \qquad \text{Eq.(1)}$$

Here "$i$" and "$j$" denote the initial and final electronic states, $n$ and $m$ denote the phonon states. Under the Born-Oppenheimer approximation, $\psi_i(r,R)$ is the $i$-th eigen state of the electronic degree of freedom for a fixed atomic configuration $R$:

$$H(r,R)\psi_i(r,R) = \varepsilon_i(R)\psi_i(r,R) \qquad \text{Eq.(2)}$$

In above equation, $R$ is just a parameter, $H(r,R)$ is the electron Hamiltonian, either in the form of many body wave function Hamiltonian, or the density functional theory (DFT). $\varepsilon_i(R)$ is the total energy of the system when the system is at the $i$-th electronic state and the atomic configuration is $R$. The phonon wave function satisfies the following equation:

$$\left[\sum_R -\frac{1}{2M_R}\nabla_R^2 + \varepsilon_i(R)\right]\phi_{i,n}(R) = E_{i,n}\phi_{i,n}(R) \qquad \text{Eq.(3)}$$

Here $M_R$ is the nuclear mass, and $E_{i,n}$ is the total energy of electron and phonon state $i,n$. The same can be said for state $\psi_j(r,R)$ and $\phi_{j,m}(R)$. In the adiabatic approximation formalism, the transition between $\Psi_{i,n}(r,R)$ and $\Psi_{j,m}(r,R)$ comes because they are not the true eigen states in the electron-phonon combined Hamiltonian due to the first and second order derivatives of $\Psi_{i,n}(r,R)$ and $\Psi_{j,m}(r,R)$ with respect to $R$, which are ignored in the Born-Oppenheimer approximation. As a result, there is a coupling between these two states:

$$\langle\Psi_{j,m}|H_{tot}|\Psi_{i,n}\rangle \approx 2\left\langle \psi_j(r,R)\phi_{j,m}(R)\left|\frac{\partial\psi_i(r,R)}{\partial R}\frac{\partial\phi_{i,n}(R)}{\partial R}\right.\right\rangle \qquad \text{Eq.(4)}$$

In writing down the above equation, we have ignored the second order derivative of $\psi_i(r,R)$ by $R$. Now, under the Frank-Condon approximation, we assume $\int \psi_j(r,R)\frac{\partial \psi_i(r,R)}{\partial R}d^{3N}r$ is independent of $R$, hence can be put out from the $R$ integration in Eq.(4), hence we have:

$$\langle\Psi_{j,m}|H_{tot}|\Psi_{i,m}\rangle \approx 2\left\langle\psi_j(r,R)\left|\frac{\partial\psi_i(r,R)}{\partial R}\right.\right\rangle\bigg|_{R=R_0}\langle\phi_{j,m}(R)|\phi_{i,n}(R)\rangle \qquad \text{Eq.(5)}$$

When the adiabatic approximation failed to yield large enough transition rate, there are many analysis for what might be wrong. A lot of blame goes to the Frank-Condon approximation from Eq.(4) to Eq.(5). High order perturbation theory to express $\psi_i(r,R)$ and $\psi_j(r,R)$ are used to show how Eq.(4) can be converted to static coupling approximation, or to show the adiabatic approximation and static coupling approximation are the same [19]. We like to express a different view here.

The Eq.(4) critically depends on the validity of Eq.(3) for all atomic configurations $R$. This includes the configuration $R$ where $\varepsilon_i(R) \approx \varepsilon_j(R)$. As we known, it is at such energy crossing point the transition happen most easily. For example, the Marcus theory can be described by a Landau-Zener transition when such energy crossing happens. However, it is well known that the Born-Oppenheimer adiabatic approximation breaks down exactly at such crossing point (much like it breaks down at conical intersection of small molecules in quantum chemistry calculations). For our problem, to state this in another way, if Eq.(2) is satisfied for every $R$, including the crossing point, as shown in Fig.1, the two valley states $i$ and $j$ are actually connected continuously by the solid line, thus they are the same state under the adiabatic approximation. As a result, there could not be an $i$ to $j$ transition in the first place. Yes, another related fact is that, one adiabatic state (the solid line) will have very fast charges with $R$ near the crossing point. As a result, the $\partial \psi_i(r,R)/\partial R$ at that $R$ will be significantly different from the derivation at $R_a$, hence the Frank-Condon approximation will break down. To avoid such fundamental problems, one has to break away from the Born-Oppenheimer adiabatic approximation of Eq.(1). In order to have an $i$ to $j$ transition, we must describe the electron wave function differently as indicated by the dotted line in Fig.1. We can call these two wave functions $\psi'_i(r,R)$ and $\psi'_j(r,R)$. They will not satisfy the Eq.(2) around the crossing point, and their transition happens mostly at the crossing point, and the dominant contribution to their coupling term $\langle \Psi'_{j,m} | H_{tot} | \Psi'_{i,n} \rangle$ also comes from the crossing point. The $\psi'_i(r,R)$ and $\psi'_j(r,R)$ of the dotted line branches in Fig.1 can be approximated by $\psi'_i(r,R) = \psi_i(r,R_a)$ and $\psi'_j(r,R) = \psi_j(r,R_a)$ as indicated by the dashed line in Fig.1 (the same $R_a$ must be used to preserve the electronic orthogonality between these two states). As a result, we have their coupling constant under the first order approximation as:

$$\langle \Psi'_{j,m}(r,R) | H_{tot}(r,R) | \Psi'_{i,n}(r,R) \rangle$$
$$= \langle \psi_j(r,R_a) \varphi_{j,m}(R) | H_a + \sum_R \frac{\partial H}{\partial R}(R - R_a) | \psi_i(r,R_a) \varphi_{i,n}(R) \rangle \quad \text{Eq.(6)}$$
$$= \sum_R \langle \psi_j(r,R_a) | \frac{\partial H}{\partial R} | \psi_i(r,R_a) \rangle \langle \varphi_{i,n}(R) | (R - R_a) | \varphi_{j,m}(R) \rangle$$

This is the static coupling formalism. Note, by approximating $\psi'_i(r,R) = \psi_i(r,R_a)$, we no longer have the $R$ derivative of $\psi_i(r,R)$ in Eq.(6), which was the cause of coupling in the adiabatic approximation. Eq.(6) captures the coupling caused by the Landau-Zener theory happen at the energy crossing point, while the coupling described by the adiabatic approximation (especially under the Frank-Condon approximation) is caused by the imperfection of the Born-Oppenheimer approximation near the valley point $R_a$. But that imperfection is rather small, hence the resulting adiabatic formula yields a too small transition rate. Thus, from this discussion, it becomes clear the static approximation should be used. Note, in both Frank-Condon approximation and static coupling approximation, there is a choice of $R_a$. One usually practice is to use the relaxed valley point position $R$ when the electron is occupying the localized state (the defect state).

## Variational electron-phonon coupling constant calculations

One key element in evaluating the nonradioactive recombination rate in most of the formalisms is the calculation of the electron-phonon coupling constants: $\langle \psi_j(r,R_a) | \partial H / \partial Q_k | \psi_i(r,R_a) \rangle$, where $Q_k$ is the k-th phonon mode coordinate. If the phonon modes are known [10], the above constants can also be calculated from $\langle \psi_j(r,R_a) | \partial H / \partial R | \psi_i(r,R_a) \rangle$. Note $H$ is the self-consistent electron Hamiltonian. In DFT, it is the SCF Kohn-Sham equation single particle Hamiltonian. Thus if $\partial H / \partial R$ needs to be calculated numerically for every $R$, there will be $3N$ SCF calculations, where $N$ is the number of atoms in the system. That makes the *ab initio* calculation extremely expensive. Recently, we have proposed a variational way to carry out the calculation for all the electron-phonon coupling constants by one additional calculation [10]. We have shown that, for a local/semilocal exchange-correlation functional, in a self-consistent Kohn-Sham calculation, if we have used $\rho_\lambda(r) = \sum_{i \in occ} |\psi_i(r)|^2 + \lambda \rho_f(r)$ for the ionic, Hartree and exchange-correlation energy evaluations, while keeping the conventional formalism for the kinetic energy and nonlocal potential, and we keep $\rho_\lambda(r)$ fixed during SCF iterations for $\psi_i(r)$, then we have:

$$\frac{d}{d\lambda} F_R = \int \rho_f(r) \frac{\partial}{\partial R} V_{tot}(r, R) d^3 r \qquad \text{Eq.(7)}$$

Here $F_R$ is the ab initio atomic force on atom $R$ calculated from Hellman-Feynman formula, and $V_{tot}$ is the self-consistent total potential in the Kohn-Sham Hamiltonian: $H(R) = -\frac{1}{2}\nabla^2 + \sum_{l,R} |\varphi_{l,R}\rangle\langle\varphi_{l,R}| + V_{tot}(r,R)$ (here $\varphi_{l,R}$ is the nonlocal potential projector for atom $R$ and angular momentum $l$). Thus

$$\left\langle \psi_j(r) \left| \frac{\partial}{\partial R} H \right| \psi_i(r) \right\rangle$$
$$= \sum_l \left\langle \psi_j(r) \left[ |\varphi_{l,R}\rangle\left\langle \frac{\partial \varphi_{l,R}}{\partial R}\right| + \left|\frac{\partial \varphi_{l,R}}{\partial R}\right\rangle\langle \varphi_{l,R}| \right] \psi_i(r) \right\rangle + \int \psi_j(r)\psi_i(r) \frac{\partial}{\partial R} V_{tot}(r,R) d^3 r \qquad \text{Eq.(8)}$$

Note, the first term is the same term as in Hellman-Feynman force evaluation, which can be calculated readily, while the second term can be calculated from Eq.(7) while using $\rho_f(r) = \psi_j(r)\psi_i(r)$.

The above formalism is derived based on local/semilocal exchange-correlation functional. Here we like to point out that the same variational approach works also for nonlocal functional like the hybrid density functional method (e.g., the screened hybrid functional of Heyd, Scuseria, and Ernzerhof (HSE) [23,24]). In our variational calculation, the total energy of the system under hybrid exchange-correlation functional will be expressed as:

$$E(\psi, R, \lambda) = -\frac{1}{2}\sum_{k \in occ}\langle \psi_k | \nabla^2 | \psi_k \rangle + \sum_{k,R,l}\langle \psi_k | \varphi_{l,R}\rangle\langle \varphi_{l,R} | \psi_k \rangle + U_\alpha(\rho_\lambda, R) -$$
$$\alpha \sum_{k,k'} \int \psi_k(r)\psi_{k'}(r)\psi_k(r')\psi_{k'}(r') v(r-r') d^3 r d^3 r' - \alpha\lambda \sum_k \int \phi_j(r)\psi_k(r)\psi_k(r')\phi_i(r') v(r-r') d^3 r d^3 r' \qquad \text{Eq.(9)}$$

Here $U_\alpha(\rho_\lambda, R)$ is the ionic, Hartree and local/semilocal exchange-correlation energy with mixing factor $\alpha$, and $\rho_\lambda$ is the charge density defined above. The $v(r-r')$ is the long range truncated Coulomb interaction kernel. When $\lambda = 0$, Eq.(9) returns to the conventional hybrid functional total energy. While carrying out SCF iteration to find the minimum energy of $E(\psi, R, \lambda)$ with respect to $\psi$, we will fix the $\rho_f(r)$, and $\varphi_i(r)$ and $\varphi_j(r)$. The corresponding single particle Hamiltonian will be the same as the conventional hybrid functional calculation except an additional exchange term: $-\alpha\lambda \int [\phi_i(r)\phi_j(r') + \phi_j(r)\phi_i(r')] v(r-r')\psi_k(r')d^3r'$. But this term will not cause any significant extra computational cost. Then following the same derivation steps as in Eq.(8) of Ref.[10], we have:

$$\frac{d}{d\lambda} F_R = \frac{d}{dR}\left[\frac{\partial E(\psi, R, \lambda)}{\partial \lambda}\bigg|_{\psi,R}\right]$$
$$= \frac{d}{dR}\left[\frac{\partial U_\alpha(\rho_\lambda, R)}{\partial \rho_\lambda}\frac{\partial \rho_\lambda}{\partial \lambda} - \alpha \sum_k \int \psi_k(r,R)\psi_k(r',R)\phi_j(r)\phi_i(r')v(r-r')d^3rd^3r'\right] \quad \text{Eq.(10)}$$
$$= \int \rho_f(r)\frac{dV_{tot}^\alpha(r,R)}{dR}d^3r - \alpha \sum_k \int \phi_j(r)\left[\frac{d\psi_k(r,R)}{dR}\psi_k(r',R) + \psi_k(r,R)\frac{d\psi_k(r',R)}{dR}\right]\phi_i(r')d^3rd^3r'$$

This is exactly the $\langle \psi_j(r)|\partial H/\partial R|\psi_i(r)\rangle$ for the hybrid functional single particle Hamiltonian when and $\rho_f(r) = \psi_i(r)\psi_j(r)$. As in the local/semilocal case, the nonlocal potential term (the first term in Eq.(8)) needs to be added. Thus, same as in the local/semilocal exchange-correlation functional case, the electron-phonon coupling constants of the hybrid functional can also be obtained by doing one additional SCF calculations of Eq.(9) with a small $\lambda$, and using Hellman-Feynman formula to calculate the atomic force $F_R(\lambda)$, then a numerical difference method will give us $dF_R/d\lambda$.

## Computational Formalisms for different methods

In this section, we will write down the actual computational formalisms to be used for different methods. For all quantum mechanical methods, the wavefunctions of a system are approximated by the product of electronic states and vibronic states as described by Eq.(1), or similar expression. The nonradiative decay probability between electron state *i* or *j* is given by the conventional Fermi "golden rule" expression:

$$W_{ij} = \frac{2\pi}{\hbar} \sum_n \sum_m p(i,n)|V_{in,jm}|^2 \delta(E_{in} - E_{jm}) \quad \text{Eq.(11)}$$

where the off-diagonal matrix elements of the electronic Hamiltonian are:

$V_{in,jm} = \langle \Psi_{j,m}(r,R)|H|\Psi_{i,n}(r,R)\rangle$

and $p(i,n)$ is the probability that the system is in the initial phonon state $\Psi_{i,n}(r,R)$, so that $\sum_n p(i,n) = 1$. Provided that the vibrational equilibrium rate considerably exceeds the nonradiative decay rate, $p(i,n)$ can be described by Boltzment distribution:

$$p(i,n) = Z^{-1} \exp(-\beta E_{in})$$

where $Z = \sum_n \exp(-\beta E_{in})$ is the partition function and $\beta = (k_B T)^{-1}$.

The off-diagonal matrix elements $V_{in,jm}$ are important in determining the nonradiative recombination coefficients. Different ways to approximate $V_{in,jm}$ constitute different formalisms. For these we have the following approximations:

### i. Static coupling theory

For the static coupling theory, $\psi_i(r,R)$ and $\psi_j(r,R)$ are replaced by $\psi_i(r,R_a)$ and $\psi_j(r,R_a)$ respectively, the off-diagonal matrix elements $V_{in,jm}$ can be expressed by Eq.(6), here we re-express it using normal phonon modes $Q_k$:

$$\begin{aligned}
&\left\langle \Psi_{j,m}(r,R) \middle| H \middle| \Psi_{i,n}(r,R) \right\rangle \\
&= \left\langle \psi_j(r,R_a)\varphi_{j,m}(R) \middle| H_a + \sum_k \frac{\partial H}{\partial Q_k}(Q_k - Q_{k,a}) \middle| \psi_i(r,R_a)\varphi_{i,n}(R) \right\rangle \\
&= \sum_k \left\langle \psi_j(r,R_a) \middle| \frac{\partial H}{\partial Q_k} \middle| \psi_i(r,R_a) \right\rangle \left\langle \varphi_{i,n}(R) \middle| (Q_k - Q_{k,a}) \middle| \varphi_{j,m}(R) \right\rangle \\
&= \sum_k C_{i,j}^k \cdot \left\langle \varphi_{i,n}(R) \middle| \mathbf{Q}_k \middle| \varphi_{j,m}(R) \right\rangle
\end{aligned} \qquad \text{Eq.(12)}$$

here $R_a$ (or $Q_{k,a}$) is the relaxed position of $i$, and by definition: $\mathbf{Q}_k \equiv Q_k - Q_{k,a}$

The relaxed state coordinations of phonon $k$ at electron states $i$ and $j$, can be calculated as:

$$\mathbf{Q}_{(i,j)k} = \frac{1}{\sqrt{M_k}} \sum_R M_R \mu_k(R) \mathbf{R}_{(i,j)}$$

Here $\mu_k(R)$ is the $k$-th phonon mode vector, and $M_R$ is the nuclear mass for atom at $R$. We adopt harmonic approximation to describe the phonon states of the initial "$i$" and final "$j$" electronic state. We have the linear relation:

$$\mathbf{Q}_j = \mathbf{Q}_i + \mathbf{K} \qquad \text{Eq.(13a)}$$

where $\mathbf{K} = Q_j(0) - Q_i(0) \equiv \Delta Q_{ji}$, or more explicitly

$$\mathbf{K}_k = \Delta Q_{ji,k} = \frac{1}{\sqrt{M_k}} \sum_R M_R \mu_k(R) \Delta R_{ji} \qquad \text{Eq.(13b)}$$

here $\Delta R_{ji} = R_j(0) - R_i(0)$ is the relaxed atomic position difference of the system at the electronic states $j$ and $i$.

Then the electron phonon coupling constant $C_{i,j}^k$ between electronic states $i$ and $j$ and phonon mode $k$ is:

$$C_{i,j}^k = \left\langle \psi_j(r, R_a) \left| \frac{\partial H}{\partial Q_k} \right| \psi_i(r, R_a) \right\rangle = \sum_R \mu_k(R) \left\langle \psi_j \left| \frac{\partial H}{\partial R} \right| \psi_i \right\rangle \qquad \text{Eq.(14)}$$

and $\sum_R M_R \mu_k(R) \mu_l(R) = \delta_{k,l}$

Now, plug in Eq.(11), we have:

$$W_{ij} = \frac{2\pi}{\hbar} \sum_{k_1,k_2} C_{i,j}^{k_1} C_{i,j}^{k_2}$$
$$\cdot \left\{ \sum_n \sum_m p(i,n) \left\langle \varphi_{j,m}(R) \left| \mathbf{Q}_{k_1} \right| \varphi_{i,n}(R) \right\rangle \left\langle \varphi_{j,m}(R) \left| \mathbf{Q}_{k_2} \right| \varphi_{i,n}(R) \right\rangle^* \delta(\hbar\omega_{jm} - \hbar\omega_{in} + \Delta E_{ij}) \right\}$$

Here $\Delta E_{ij}$ is energy difference of the initial and final electronic states after atomic relaxation with no phonon contributions while $\hbar\omega_{in}$ and $\hbar\omega_{jm}$ denote the phonon energies in the initial and final states (with multiple phonon modes and multiple phonons). Using the Dirac distribution function,

$$\delta(\omega) = \frac{1}{2\pi} \int_{-\infty}^{\infty} e^{i\omega t} dt$$

We can get,

$$W_{ij} \equiv \frac{2\pi}{\hbar} \sum_{k_1,k_2} C_{i,j}^{k_1} C_{i,j}^{k_2} \cdot A_{ij}^{k_1 k_2} \qquad \text{Eq.(15)}$$

where: $A_{ij}^{k_1 k_2} = \frac{1}{2\pi Z} \int_{-\infty}^{\infty} \chi_{ij}^{k_1 k_2}(t,T) e^{-it\frac{\Delta E_{ij}}{\hbar}} dt$, $\qquad \text{Eq.(16)}$

$$\chi_{ij}^{k_1 k_2}(t,T) = \text{Tr}\left[ \mathbf{Q}_{k_1} e^{-itH_j/\hbar} \mathbf{Q}_{k_2} e^{-(\beta - it)/H_i/\hbar} \right] \qquad \text{Eq.(17)}$$

This is the Herzberg-Teller (HT) terms, where $H_i$ and $H_j$ are the phonon Hamiltonians for the initial and final electronic states, where $\beta = (k_B T)^{-1}$. The indices $k_1$, $k_2$ run over the $N_{vib}$ normal coordinates, where $N_{vib}$ is the number of vibrational degrees of freedom. The trace in Eq.(17) can be computed on the basis of $\mathbf{Q}$, as follows,

$$\chi_{ij}^{k_1 k_2}(t,T) = \int_{-\infty}^{\infty} d\mathbf{Q} \langle \mathbf{Q} | \mathbf{Q}_{k_1} e^{-itH_j/\hbar} \mathbf{Q}_{k_2} e^{-(\beta-it)/H_i/\hbar} | \mathbf{Q} \rangle \quad \text{Eq.(18)}$$

This term can be evaluated using path integral techniques and Gaussian integration, and the derivation presented by Borrelli et al. [25-27] is very efficient which has been implemented here. Let us introduce the following diagonal ($N_{vib} \times N_{vib}$) matrices, $a(\tau_i)$, $a(\tau_j)$, $c(\tau_i)$, $c(\tau_j)$, $d(\tau_i)$, $d(\tau_j)$, whose $k$th diagonal terms are given by ($\xi = i, j$),

$$a(\tau_\xi)_k = \frac{\omega_k}{\sinh(i\hbar\omega_k \tau_\xi)} \quad \text{Eq.(19a)}$$

$$c(\tau_\xi)_k = \omega_k \coth(i\hbar\omega_k \tau_\xi / 2) \quad \text{Eq.(19b)}$$

$$d(\tau_\xi)_k = \omega_k \tanh(i\hbar\omega_k \tau_\xi / 2) \quad \text{Eq.(19c)}$$

where $\tau_i = -t - i\beta$ and $\tau_j = t$, we have used the assumption that the phonon mode in $i$ and $j$ states are the same $k_1 = k_2 = k$ and $\omega_k$ is the frequency of the $k$th harmonic oscillator. From the above matrices, the following diagonal matrices can be defined,

$$C(\tau_i, \tau_j) = c(\tau_j) + c(\tau_i) = \omega_k \left( \coth\left(\frac{i\hbar\omega_k \tau_j}{2}\right) + \coth\left(\frac{i\hbar\omega_k \tau_i}{2}\right) \right) \quad \text{Eq.(20a)}$$

$$D(\tau_i, \tau_j) = d(\tau_j) + d(\tau_i) = \omega_k \left( \tanh\left(\frac{i\hbar\omega_k \tau_j}{2}\right) + \tanh\left(\frac{i\hbar\omega_k \tau_i}{2}\right) \right) \quad \text{Eq.(20b)}$$

To evaluate the HT term of Eq.(17), a $N_{vib}$–dimensional column vector, $D_{HT}$, and a ($N_{vib} \times N_{vib}$) matrix, $A_{HT}$, defined by Borrelli et al. [25,27] are also needed (with $\mathbf{K}$ defined in Eq.(13a,b)),

$$D_{HT} = -D^{-1} d(\tau_j) \mathbf{K} \quad \text{Eq.(21a)}$$

$$A_{HT} = \frac{1}{2}\left(D^{-1} - C^{-1}\right) + D_{HT}\left(D_{HT}\right)^T \quad \text{Eq.(21b)}$$

Under the assumption that the phonon modes in $i$ and $j$ states are the same, we have $\chi_{ij}^{k_1 k_2} = \chi_{ij}^{k_1} \delta_{k_1 k_2}$, then [25, 26]:

$$\chi_{ij}^{k}(t,T) = \sqrt{\frac{\det(a(\tau_j))\det(a(\tau_i))}{(i\hbar)^{2N} \det(C)\det(D)}} \times \exp\left[-\mathbf{K}^T d(\tau_j)\mathbf{K} + \mathbf{K}^T d(\tau_j) D^{-1} d(\tau_j)\mathbf{K}\right] (A_{HT})_{kk} \quad \text{Eq.(22)}$$

Introducing Eq.(22) in Eq.(16), we can numerically carry out the time integration to get $A_{ij}^{k_1 k_2} = A_{ij}^{k_1} \delta_{k_1 k_2}$. The time integration in Eq.(16) converges well especially when a small damping term for the limits of $-\infty$ and $\infty$ is used.

### ii. Adiabatic approximation

The formalism for the adiabatic approximation has been given in our previous work, especially Eq.(2) of Ref.[10]. As above, we have also assumed that the phonon modes at electronic states *i* and *j* are the same except a shift in the origin. We have also used the strong coupling limit formula [14], which has been shown to be valid in the case of nonradiative decay. As a result of this limit, the time integration (similar to Eq.(16) above) can be evaluated out analytically. Thus at the end, all the expressions are analytical, the nonradiative rates can be calculated as long as the phonon modes and electron-phonon coupling constants are known.

### iii. Classical Marcus theory

The transfer rate in Marcus theory is expressed as [21]:

$$\tau^{-1} = |V_c|^2 \sqrt{\frac{\pi}{\lambda k_B T \hbar^2}} \exp\left[-\left(\lambda + E_j - E_i\right)^2 / 4\lambda k_B T\right] \qquad \text{Eq.(23)}$$

$V_c$ is the electronic coupling between electronic states *i* and *j*, $\lambda$ and is the reorganization energy of the system (the atomic relaxation energy after the electron transferred from *i* to *j*), $E_i$ and $E_j$ represent the total energy of the system (after atomic relaxation) at electronic state *i* and *j*, as shown in Figure 2. Note that these are the energies at the minimum of the energy valley in the atomic (phononic) degree of freedom. Thus they do not include the phonon energies. While $E_i$, $E_j$ and $\lambda$ are well defined, one major challenge is to calculate $V_c$. One common approach is to use diabatic states to represent *i* and *j* electron wave functions. They are not the eigen states, thus can have coupling under an fixed electron Hamiltonian at $R_a$. One way is to construct a maximally localized defect state using a mixture of the eigenstates $\psi'_j = c_1 \psi_i + c_2 \psi_j$, and set the conduction band state $\psi'_i$ to be orthogonal to this defect diabatic state. Then $V_c = \langle \psi'_i | H | \psi'_j \rangle$. But we found that the resulting coupling constant is rather small. Thus we conclude the coupling cannot be caused by diabatic states.

Another approach is to use an external potential to perturb the eigen energies $\varepsilon_i$ and $\varepsilon_j$ until they cross each other, then the anticrossing energy gap should be $2V_c$ [21]. Unfortunately, in our case, the perturbation potential also causes coupling by itself, as a result, different ways of doing perturbation can yield very different results. Physically, however, the Marcus theory does describe a Landau-Zener transition when the electronic eigen energies of $i$ and $j$ cross each other caused by the phonon fluctuation. Thus, instead of using arbitrary perturbations, we can use the actual phonon distortion. Since Marcus theory can be considered as a 1D theory, one nature approach is to perturb the system along the transition degree of freedom, defined as: $\Delta R = R_j - R_i$. If we distort the system large enough along this direction, $\varepsilon_i$ and $\varepsilon_j$ can cross each other, then the coupling between $\psi_i$ and $\psi_j$ when they cross can be used for $V_c$. One way is to actually do the displacement along $\Delta R$, and carry out *ab initio* calculation, and get the anticrossing of $\varepsilon_i$ and $\varepsilon_j$. But we found that the required distortion is so large, the electronic structure has been totally changed, and we can no longer identify $\varepsilon_i$ and $\varepsilon_j$. As a result, we have used an analytical expression to calculate the coupling:

$$V_c = \langle \psi_i(Q_i) | H(Q_c) | \psi_j(Q_i) \rangle = \langle \psi_i | \partial H / \partial Q | \psi_j \rangle (Q_c - Q_i) \qquad \text{Eq.(24)}$$

Here $Q$ is the 1D degree of freedom along $\Delta R$, and $Q_i$ is the initial position at state $i$, and $Q_c$ is the crossing point when $E_i$ and $E_j$ cross each other, which can be estimated using parabolic approximation of $E_i(Q)$ and $E_j(Q)$. Note, in using the above formula, we can say that the coupling $V_c$ is not caused by diabatic coupling at $H(Q_i)$ ($\langle \psi_i(Q_i) | H(Q_i) | \psi_j(Q_i) \rangle = 0$) instead, it is induced by phonon modes along $\Delta R$.

iv. **Quantum CT theory**

The above classical Marcus theory can be re-derived quantum mechanically for the phonon degree of freedom while keeping the coupling constant $V_c$ as the same. The quantum mechanical formula was derived by Jortner [28] and Lin et al [29]. It starts with the Fermi golden rule:

$$k_{CT} = \frac{2\pi}{\hbar^2}|V_c|^2 \sum_{n,m} p(i,n)\left|\langle \varphi_{j,m} | \varphi_{i,n}\rangle\right|^2 \delta(E_{in} - E_{jm}) \qquad \text{Eq.(25)}$$

where $V_c$ is the electronic coupling between initial and finial states, $\varphi$ corresponds to vibrational wavefunctions. By using Slater sum (Mehler's formula), Nan and et al.[16] have derived the quantum CT rate expressed as:

$$k_{CT} = \frac{|V_c|^2}{\hbar^2}\int_{-\infty}^{\infty} dt \exp\left\{i\omega_{fi} t - \sum_k S_k\left[(2\bar{n}_k + 1) - \bar{n}_k e^{-i\omega_k t} - (\bar{n}_k + 1)e^{i\omega_k t}\right]\right\} \qquad \text{Eq.(26)}$$

where $\bar{n}_k = \frac{1}{e^{\hbar\omega_k/k_B T} - 1}$ denotes the population of $k$-th normal mode and $\omega_k$ is its frequency. $S_k = \frac{\lambda_k}{\hbar\omega_k} = \frac{1}{2}\hbar^{-1}\omega_k(\Delta Q_k)^2$ is the Huang-Rhys factor measuring the charge-phonon coupling strength. The above auxiliary time integral can be carried out numerically without much difficulty (it converges well when a slowly damping factor is used towards $\infty$ and $-\infty$).

### v. 1D quantum formula

In the work of A. Alkauskas et al., a quantum mechanical 1D model is presented [17]. In this 1D model, as discussed above, the atomic movement is along the $\Delta R = R_j - R_i$ direction, and one phonon degree of freedom along this direction is considered. Numerically, the coordination $Q$ along this direction is defined as:

$$Q^2 = \sum_\alpha M_\alpha (R_\alpha - R_{i:\alpha})^2$$

Here, $\alpha$ is the atom index, $M_\alpha$ is the nuclear mass. In the 1D model, only a single electron-phonon coupling matrix element:

$$W_{ij} = \left\langle \Psi_i \left| \frac{\partial H}{\partial Q} \right| \Psi_j \right\rangle$$

is needed. It can be calculated numerically. The capture coefficient is calculated using the static coupling model as:

$$C_p = \frac{2\pi}{\hbar^2} W_{ij}^2 \sum_n p(i,n) \sum_m \left|\langle \varphi_{j,m} | Q - Q_i | \varphi_{i,n}\rangle\right|^2 \delta(\Delta E_{ij} + m\hbar\Omega_i - n\hbar\Omega_j) \qquad \text{Eq.(27)}$$

Here, $\Omega_i$ and $\Omega_j$ are the effect phonon modes along the transition degree of freedom at electronic state $i$ and $j$. In our approximation, we have set $\Omega_i = \Omega_j$. Note here, only one effective phonon mode is considered, hence $\varphi_{i,n}$ and $\varphi_{j,m}$ are single phonon mode wave functions. In Ref.[17], how Eq.(27) is evaluated was not discussed in detail. Here, we have used the same formalism as in the multiphonon case. For only one effective phonon mode, we only need to set $k_1 = k_2 = 1$ to change Eq.(15) to calculate the above equation. In numerically calculating the time integral of Eq.(16), a relatively large damping factor is used to yield a smooth transition rate as a function of $\Delta E_{ij}$, this is equivalent of smearing the delta function in Eq.(27).

Note, one difference between Eq.(27) and Eq.(25) is that, in Eq.(25), the ($Q - Q_i$) factor is pull out as a fixed factor ($Q_c - Q_i$), which is combined with $W_{ij}$ to give $V_C$. The fact to pull out $Q - Q_i$ is like a classical approximation, which represents its effect by the crossing point value. On the other hand, in Eq.(25), all the phonon modes are used, while in Eq.(27), only one single phonon mode is used. However, if we try to include all the phonon modes in Eq.(27), besides the effective phonon mode along the $\Delta R$ direction, all the other phonon modes $k$ must be orthogonal to this effective phonon mode, as a result, their $\Delta Q_k = Q_k(j) - Q_k(i)$ are all zero. Then the corresponding $\langle \varphi_{i,n}(k) | \varphi_{j,m}(k) \rangle = \delta_{n,m}$, thus there is no effect for all the other phonon modes. In another word, the Eq.(27) is the same if all the phonon modes are included.

## Results:

In the following, we present the results of the two complex defects GaP:$Zn_{Ga}$-$O_P$ and GaN:$Zn_{Ga}$-$V_N$ using different methods. The bottleneck for the nonradiative recombination is the charge transfer from the minority carrier to the defect state. For GaP:$Zn_{Ga}$-$O_P$, this is from the electron in the conduction band to the defect state; for GaN:$Zn_{Ga}$-$V_N$, this is from the hole in the valence band to the defect state.

### A. GaP:$Zn_{Ga}$-$O_P$

1. Atomic structure and formation energy

A 64-atom ($2 \times 2 \times 2$) supercell is used to calculate the formation energy of the $Zn_{Ga}$-$O_P$ in GaP using the screened hybrid functional of HSE [23,24] and generalized gradient approximation (GGA) atomic relaxation. The cutoff energy for the plane-wave basis is 400 eV. The mixing parameter of HSE was set to 0.2. The GGA calculated equilibrium lattice parameters of GaP, a=5.49 Å. This part of the calculation uses the commercial code VASP (the Vienna *ab initio* simulation package) [30].

The atom structures for $Zn_{Ga}$-$O_P$ in GaP at different charge states are shown in Fig.3. The impurity atoms Zn and O drift from the equilibrium positions of GaP bulk. For different charge states, the impurity atoms are also drift. When one electron is captured, the bond length between Zn and O changes from 3.16 Å of $(Zn_{Ga}$-$O_P)^0$ to 2.42Å of $(Zn_{Ga}$-$O_P)^-$.

Following Ref.[31], the formation energy $\Delta E^q$ of $Zn_{Ga}$-$O_P$ defect with charge $q$ is calculated as shown in Fig. 4. Our HSE calculated 0/- transition energy is at 2.023 eV above the VBM as shown in Fig.4. The $Zn_{Ga}$-$O_P$ center in GaP is a neutral center for which both the recombination luminescence and electron-capture cross section are accurately known experimentally. The experimentally established energy of the exciton bound to the $Zn_{Ga}$-$O_P$ center at helium temperature is approximately 2.02 eV. Henry et al. found the binding energy of the the defect bounded exciton to be approximately 0.037 eV [22]. Adding this binding energy to the bounded exciton energy, we get the defect level at about 2.057 eV above the VBM. This is close to our calculated 2.023 eV result. Thus the transition energy from the CBM to the defect level is $\Delta E$ =0.282 eV for a 2.34 eV GaP band gap (by definition, this is the energy the system lowers by transferring an minority electron carrier from the conduction band to the defect state. It is thus the $\Delta E_{ij}$ in Eq.(16) or all the other equations). The relevant electron-phonon coupling should be between the impurity states and the conduction band states, which can be represented by the CBM state. In Ref.[17], a prefactor is used to represent the wave function amplitude difference near the defect between the CBM state and an actual itinerate state at the finite temperature. However, for neutral defect, that factor is rather close to 1 (e.g., 1.02), thus we have ignored that prefactor in the following discussions.

2. Impurity wavefunction and phonon DOSs

For the nonradiativecarrier recombination process of GaP:$Zn_{Ga}$-$O_P$, the relevant electron-phonon coupling should be between the impurity state and the conduction band states, which can be represented by the CBM state. The impurity state wavefunction is localized in a 64-atom supercell as illustrated in Fig. 5.

Then we used a "combined dynamic matrix"(CDM) method as described in Ref.[10] to calculate the phonon modes and phonon DOSs are show in Fig. 6. The CDM requires only the calculation of atomic displacements for a few atoms within $R_c$ distance of the defect center to get the phonon mode of the whole system. We used $R_c$ (=6.0 a.u.), for which the CDM describes well the impurity DOS [10]. There are some localized phonon mode peaks inside the gap of bulk phonon DOS. We also see some significant shift for the acoustic modes. For every phonon mode, we get the frequency $\omega_k$ and eigenvector $\mu_k(R)$ with a normalization of $\sum_R M_R \mu_{k_1}(R)\mu_{k_2}(R) = \delta_{k_1,k_2}$. We also get the displacement $\mathbf{K}_k$ from the atomic displacement $R_i - R_j$ after the electron charge transfer by Eq.(13b).

3. Configuration coordinate diagram

Based on the 1D quantum formula method, we calculate the 1D cc diagram for $Zn_{Ga}$-$O_P$ center in GaP as shown in Fig. 7. The total energy points at different displacement Q along the transition degree of freedom are actually calculated, and shown in Fig.7 for the neutral and "-" charged state.

When Q<7 amu$^{1/2}$ Å, the solid curves present parabolic fit to the energy values. But when Q>7 amu$^{1/2}$ Å, there seem to have some sudden change for its electronic structure, and which result in a sudden change for the total energy. Thus, in line with the harmonic approximations for the phonon modes in all our derivations, we have fitted only the data points of Q<7 amu$^{1/2}$ Å with parabolic curves (the blue and green lines in Fig.7) to predict their behaviors. For $(Zn_{Ga}$-$O_P)^0$, the equilibrium configuration coordinate is at Q=0 amu$^{1/2}$ Å; for $(Zn_{Ga}$-$O_P)^-$, the equilibrium configuration coordinate is at Q=4.43 amu$^{1/2}$ Å. As shown in Fig. 7, the fitting lines of two states cross at $Q_c$=19.09 amu$^{1/2}$ Å.

4. Calculated capture coefficients

We have used the Eq. (3), (8) and (9) in our previous paper [10] to calculate the electron-phonon coupling constant by two SCF calculations. Thus, we get the $C_{i,j}^k$, $\omega_k$, $\mathbf{K}_k$ of every phonon mode $k$ and $\Delta E$. From these, we calculated the nonradiative decay probability $W_{ij}$ using different formulas. The capture rate constant is $B_n = W_{ij} \cdot V$ ($V$ is the volume of the supercell) and the capture cross section $\sigma_n = B_n / v_n$ ($v_n$ is the mean thermal velocity of the electron) can also be calculated. At T=300 K, we get $B_n = 4.30 \times 10^{-8}$ cm$^3$/s and $\sigma_n = 2.15 \times 10^{-15}$ cm$^2$ by static coupling theory, which agrees well with the experimental result $\sigma_n = \left(2_{-1}^{+2}\right) \times 10^{-15}$ cm$^2$. On the other hand, if the adiabatic approximation is used, the calculated capture coefficient is only $B_n = 3.32 \times 10^{-10}$ cm$^3$/s, two orders of magnitudes smaller than the static coupling theory.

**B. For the 1D-(1-Phonon) formula**

The electron-phonon coupling constant $W_{ij} = \left\langle \psi_i \left| \frac{\partial H}{\partial Q} \right| \psi_j \right\rangle$ is calculated as

$$W_{ij} = \sum_R \left\langle \psi_i \left| \frac{\partial H}{\partial R} \right| \psi_j \right\rangle \cdot \frac{R_c - R_i}{Q_c - Q_i} = 0.25 \times 10^{-2} \text{ eV/ amu}^{1/2} \text{ Å} \qquad \text{Eq.(28)}$$

with $\left\langle \psi_i \left| \frac{\partial H}{\partial R} \right| \psi_j \right\rangle$ being calculated with the variational method. Alternatively, one can also directly calculate the above $W_{ij}$ by making a numerical displacement along the transition direction, and calculate $\left\langle \psi_i | \Delta H | \psi_j \right\rangle$ numerically (one can obtain this quantity by calculating $(\varepsilon_j - \varepsilon_i) \left\langle \psi_i(R_a) | \psi_j(R_a + \Delta R) \right\rangle$). We found the result is almost the same as from Eq.(28), e.g, for GaP:Zn$_{Ga}$-O$_P$ charged state $W_{ij} = (\varepsilon_j - \varepsilon_i) \left\langle \psi_i | \frac{\partial \psi_j}{\partial Q} \right\rangle = 0.22 \times 10^{-2}$ eV/ amu$^{1/2}$ Å. The other calculated variables needed in the 1D formula are shown in Tab. I. g is the degeneracy factor of the initial state; it reflects the fact that there might exist a few equivalent energy-degenerate (or nearly degenerate) levels of the initial state. The resulting nonradiatice capture coefficient is $B_n = 1.68 \times 10^{-10}$ cm$^3$/s. It is two orders of magnitude smaller than the multi-phonon static approximation result. This is expected as discussed in Ref.[17], since only one phonon mode is used, and the phonon mode which induces the electron-phonon coupling is forced to be the same as the phonon modes which involved in the energy conservation. Such restriction can reduce the transition rate.

TABLE I. The key variables calculated for the 1D quantum formula method

|  | $W_{ij}$ (eV/amu$^{1/2}$Å) | $|Q_i - Q_j|$ (amu$^{1/2}$Å) | $\hbar\Omega_j$ (meV) | $S_j$ | $V_C$ (eV) | Volume (Å$^3$) | $\Delta E_{ij}$ (eV) | $\lambda$ (eV) | $g$ |
|---|---|---|---|---|---|---|---|---|---|
| GaP:Zn$_{Ga}$-O$_P$ | $0.25\times10^{-2}$ | 4.43 | 5.38 | 12.61 | 0.11 | 1326 | 0.282 | 0.19 | 4 |

For Marcus theory and quantum CT rate, the electronic coupling $V_c$ is also calculated by $C_{i,j}^k$ and $Q_c$, as:

$$V_c = \left\langle \psi_i \left| \frac{\partial H}{\partial Q} \right| \psi_j \right\rangle * (Q_c - Q_i) = 0.48 \text{eV}$$

This is related to $W_{ij}$ in Eq.[30] as: $V_c = W_{ij}(Q_c - Q_i) = 0.25\times10^{-2} \times 19.09 = 0.048$ eV. The $\lambda$ in the Marcus theory is 0.19eV. Based on this electronic coupling $V_c$ and all phonon modes, we calculated the capture coefficients are $7.32\times10^{-8}$ cm$^3$/s by Marcus theory and $6.44\times10^{-8}$ cm$^3$/s by quantum CT rate at room temperature. In these cases, they are relatively close to the experimental values and the multi-phonon static approximation results.

The capture coefficients calculated by different methods are summarized in Table II. The results by static coupling theory are most consistent with the experiment results.

TABLE II. The capture coefficients (cm$^3$/s) by different methods at T=300K. The 1D quantum formula result for GaN:Zn$_{Ga}$-V$_N$ from Ref.[17] is also listed as we have failed to reproduce their result.

|  | Exp | Static | Adiabatic | Marcus theory | Quantum CT rate | 1D quantum formula |
|---|---|---|---|---|---|---|
| GaP:Zn$_{Ga}$-O$_P$ | $\left(4_{-1}^{+2}\right)\times10^{-8}$ [8] | $4.30\times10^{-8}$ | $3.32\times10^{-10}$ | $7.32\times10^{-8}$ | $6.44\times10^{-8}$ | $1.68\times10^{-10}$ |
| GaN:Zn$_{Ga}$-V$_N$ | $3.0\times10^{-7}$ [18] | $1.46\times10^{-7}$ | $5.57\times10^{-10}$ [10] | $1.18\times10^{-8}$ | $1.21\times10^{-8}$ | $1.5\times10^{-9}$ <br> $1.0\times10^{-8}$ [17] |

We also calculated the capture coefficients as functions of $T$ for GaP:$Zn_{Ga}$-$O_P$ under different methods. As shown in Fig 8, the capture coefficients increase with $T$ under the static, adiabatic and 1D quantum formula; and the capture coefficients have a peak under Marcus theory and quantum mechanical CT rate. For the 1D formula, its rate drops much faster than the multi-phonon formulas of adiabatic and static approximations. One possible reason is that, at lower temperature, the 1D formula will be more difficult to satisfy the energy conservation by the single phonon mode (mostly an optical phonon mode).

The comparison between the experiment results [8] and our calculation results of GaP:$Zn_{Ga}$-$O_P$ for their temperature dependence are shown in Fig. 9. The calculation results by static coupling theory are within the experimental uncertainty [8], although only two temperatures exist from the experiments.

Fig 10, shows the capture coefficients as a function of the defect state binding energy $\Delta E$ for different formulas. To calculate this, we let $\Delta E$ change while keeping all the other parameters (coupling constants, etc) in all the formalisms. Besides the adiabatic formula, all the results have a peak at about $\Delta E$ =0.2 eV (the 1D quantum formula has a peak around $\Delta E$ =0.15 eV). In Marcus theory, this peak appears at $\Delta E - \lambda = 0$. We note that, in this case, at room temperature, the multi-phonon static formula, the Marcus theory, and the CT formula give rather similar results.

## C. GaN:$Zn_{Ga}$-$V_N$

The detail of GaN:$Zn_{Ga}$-$V_N$ calculation have been reported in our previous paper [10], which used adiabatic approximation to calculate the capture coefficients. Now, we can use the phonon modes and electron phonon coupling constant $C_{i,j}^k$ to calculate capture coefficients by all the other methods, including the static coupling theory. The 1D method has already been calculated by Alkauskas and et al. for this system [17]. Here we report the results of all 5 different methods for a comparative study in Table. II. Note, for the 1D method, in Ref.[17], there is no details for how they get the final result. We cannot reproduce the result in Ref.[17] even if we take the parameters from Ref.[17]. This result is also shown in Table.II. Our result is about a factor of $2\pi$ smaller than the result from Ref.[17].

We also calculated the capture coefficients as functions of T for GaN:$Zn_{Ga}$-$V_N$, and the results are shown in Fig 11. At the temperature range we have studied, the capture coefficients calculated by static coupling theory are the highest and the results by adiabatic approximation are the lowest. At high temperature, the results by Marcus theory and quantum CT rate are close, and the results by 1D quantum formula are close to the results by adiabatic approximation. Similar to the case of GaP:$Zn_{Ga}$-$O_P$, the temperature dependences for the two multi-phonon formalisms: the static coupling theory and adiabatic approximation theory, are rather similar, they both decay monotonically with decreasing temperature. On the other hand, in this case, the Marcus theory and quantum CT rate increase with decreasing temperature, while the 1D formula results have a maximum rate as a relatively high temperature.

In our previous paper [10], we calculated the nonradiative recombination capture coefficients of GaN:$Zn_{Ga}$-$V_N$ using adiabatic formulation, the result is $5.57 \times 10^{-10}$ cm$^3$/s. In this paper, we use the same atomic structures and electron phonon coupling constant $C_{ij}^k$ to calculate the capture coefficients by static coupling theory, the result is $1.46 \times 10^{-7}$ cm$^3$/s, which is more than two order of magnitude larger than the adiabatic result, and it is close to the experiment results $B_p = 3 \times 10^{-7}$ to $3 \times 10^{-6}$ cm$^3$/s [18]. We note that, for this system, the Marcus theory and quantum CT theory yield an order of magnitude too small capture coefficients as shown in Table.II.

**D. Marcus theory and Quantum CT theory of nonradiative transitions**

In the strong coupling ($\sum_j S_j \gg 1$) and high-temperature limits ($\hbar\omega_j / k_B T \ll 1$, $\bar{n}_j = k_B T / \hbar\omega_j$), Eq.[26] can be integrated, and the Quantum CT formula becomes Marcus formula. We plot the capture coefficients of these two methods as functions of temperature in Fig. 12.

As we can see, at high-temperature, the results are the same between Marcus theory and Quantum CT rate. When the temperature decreasing, the capture coefficients increase firstly and then decrease. For GaP: $Zn_{Ga}$-$O_P$, the peak value of capture coefficient is $6.47\times10^{-8}$ cm$^3$/s at T=350 K by Quantum CT rate and $7.36\times10^{-8}$ cm$^3$/s at T=260 K by Marcus theory. For GaN:$Zn_{Ga}$-$V_N$, the peak value of capture coefficient is $1.18\times10^{-8}$ cm$^3$/s at T=280 K by Marcus theory and no peak value by Quantum CT rate. At very low temperature, the Quantum CT result is always bigger than the Marcus theory result. This is because of the quantum tunneling effects presented in the quantum CT result, but not in the Marcus theory.

In both of Marcus theory and Quantum CT rate, the electron phonon coupling is very important parameter, which is calculated by *ab initio* density functional theory in this work. As shown in Tab. II, for GaN:$Zn_{Ga}$-$V_N$ the results by Marcus theory and Quantum CT rate can be an order of magnitude smaller than the results by static coupling theory.

**E. Static and 1D quantum formula of nonradiative transitions**

The 1D quantum formula is based on the static coupling theory and it use only one special phonon mode to replaces the sum over all real phonon modes. For GaN:$Zn_{Ga}$-$V_N$, the capture coefficients calculated by 1D quantum formula are larger than the results by adiabatic formulation as shown in Tab. II. But when compared with the experiments, the capture coefficients calculated by 1D quantum formula are still significantly smaller than the experimental results as shown in Tab. II.

## Conclusions

In this paper, we use static coupling theory, adiabatic approximation, Marcus theory, quantum CT theory and 1D quantum formula to calculate the capture coefficients of two complex defects GaP:$Zn_{Ga}$-$O_P$ and GaN:$Zn_{Ga}$-$V_N$. Comparing the different methods, the results by the static coupling theory are most consistent with experiment. In our opinion, the static coupling theory is intrinsically more appropriate in describing the nonradiative transition in our systems. This is because it describes correctly the transition electron wave functions at the atomic configuration when their energies cross each other. In our study, all the parameters, including atomic structure, formation energy and electron-phonon coupling constants, have been calculated by *ab initio* density functional theory. All the phonon modes are considered. More specifically, we have the following results: (1) Five different formalisms are investigated, and we found the static coupling theory is the best when compared with experiments; (2) The detail formulas of these five formalisms are presented; (3) A proof is presented for how the variational method to calculate the electron-phonon coupling can also be used for hybrid functions; (4) The Marcus theory and quantum CT theory always give similar results; (5) While for GaP:$Zn_{Ga}$-$O_P$, the Marcus theory and quantum CT theory give larger capture coefficients than the multi-phonon static theory, in GaN:$Zn_{Ga}$-$V_N$, their results are an order of magnitude smaller than the multi-phonon static coupling result; (6) the 1D quantum formula gives results typically bigger than the adiabatic coupling results, but smaller than the multi-phonon static coupling results; (7) the temperature dependences of the multi-phonon static and adiabatic coupling results are similar, and between Marcus theory and quantum CT are also similar, but the temperature dependences between these two groups, and the 1D quantum formula results are all very different.

## Acknowledgement

We like to thank Dr. A. Alkauskas for discussions. Shi is supported by the National Natural Science Foundation of China under (Grant No. 11374328, 61325022, 11327804), the National Basic Research Program of China (973 Program No. 2012CB619305), and the Foundation of CNIC, CAS under Grant No. XXH12503-02-03-2(03).Wang is supported through the Theory of Material project by the Director, Office of Science (SC), Basic Energy Science (BES)/Materials Science and Engineering Division (MSED) of the U.S. Department of Energy (DOE) under the contract No. DE-AC02-05CH11231. It uses the resources of National Energy Research Scientific Computing center (NERSC) supported by the U.S. Department of Energy.We are grateful for the professional services offered by the Platforms of Characterization & Test at Suzhou Institute of Nano-Tech and Nano-Bionics (SINANO), and Supercomputing Center, CNIC, Chinese Academy of Science.

## References:

[1]   W. Shockley and W. T. Read, Phys. Rev. **87**, 835 (1952).

[2]   R. N. Hall, Phys. Rev. **87**, 387 (1952).

[3]   J. Nelson, The Physics of Solar Cells (Imperial College, London, 2003).

[4] A. M. Stoneham, Rep. Prog. Phys. **44**, 79 (1981).

[5] T. Tsuchiya, Appl. Phys. Express **4**, 094104 (2011).

[6] M. A. Reshchikov, A. A. Kvasov, M. F. Bishop, T. McMullen, A. Usikov, V. Soukhoveev, and V. A. Dmitriev, Phys. Rev. B **84**, 075212 (2011).

[7] S. Jursenas, S. Miasojedovas, G. Kurilcik, A. Zukauskas, and P. R. Hageman, Appl. Phys. Lett. **83**, 66 (2003).

[8] C. H. Henry and D. V. Lang, Phys. Rev. B **15**, 989 (1977).

[9] R. Passler, Czech. J. Phys. B **39**, 155 (1989).

[10] L. Shi and L. W. Wang, Phys. Rev. Lett. **109**, 245501 (2012).

[11] G. Helmis, Ann. Phys. (Lpz.) **19**, 41 (1956).

[12] R. Passler, Czech. J. Phys. **24**, 322 (1974).

[13] R. Passler, Czech. J. Phys. **32**, 846 (1982).

[14] K. F. Freed and J. Jortner, J. Chem. Phys. **52**, 6272 (1970).

[15] R. A. Marcus. Rev. Mod. Phys. **65**, 599 (1993).

[16] Nan, X. Yang, L. Wang, Z. Shuai, and Y. Zhao, Phys. Rev. B **79**, 115203 (2009).

[17] A. Alkauskas, Q. Yan and C. G. Van de Walle, Phys. Rev. B **90**, 075202 (2014).

[18] M. A. Reshchikov, A. J. Olsen, M. F. Bishop and T. McMullen, Phys. Rev. B **88**, 075204 (2013).

[19] K. Huang, ScientiaSinica **24**, 27 (1981).

[20] K. P. McKenna and J. Blumberger, Phys. Rev. B **86**, 245110 (2012).

[21] K. Tarafder, Y. Surendranath, J. H. Olshansky, A. P. Alivisatos, and L. W. Wang, J. Am. Chem. Soc. **136**, 5121 (2014).

[22] J. S. Jayson, R. Z. Bachrach, P. D. Dapkus, and N. E. Schumaker, Phys. Rev. B **6**, 2357 (1972).

[23] J. Heyd, G. E. Scuseria, and M. Ernzerhof, J. Chem. Phys. **118**, 8207 (2003).

[24] J. Heyd, G. E. Scuseria, and M. Ernzerhof, J. Chem. Phys. **124**, 219906 (2006).

[25] F. J. A. Ferrer, J. Cerezo, J. Soto, R. Improta, and F. Santoro, Comput. Theor. Chem. **1040-1041**, 328 (2014).

[26] A. Baiardi, J. Bloino, and V. Barone, J. Chem. Theory Comput. **9**, 4097 (2013).

[27] R. Borrelli, A. Capobianco, and A. Peluso, J. Phys. Chem. A **116**, 9934 (2012).

[28] J. Jortner, J. Chem. Phys. **64**, 4860 (1976).

[29] S. H. Lin, C. H. Chang, K. K. Liang, R. Chang, Y. J. Shiu, J. M.Zhang, T. S. Yang, M. Hayashi, and F. C. Hsu, Adv. Chem. Phys. **121**, 1 (2002).

[30] Kresse and J. Hafner, Phys. Rev. B 47, 558 (1993); 49, 14251 (1994); G. Kresse and J. Furthmuller, Phys. Rev. B **54**, 11169 (1996).

[31] S.-H. Wei, Comput. Mater. Sci. **30**, 337 (2004).

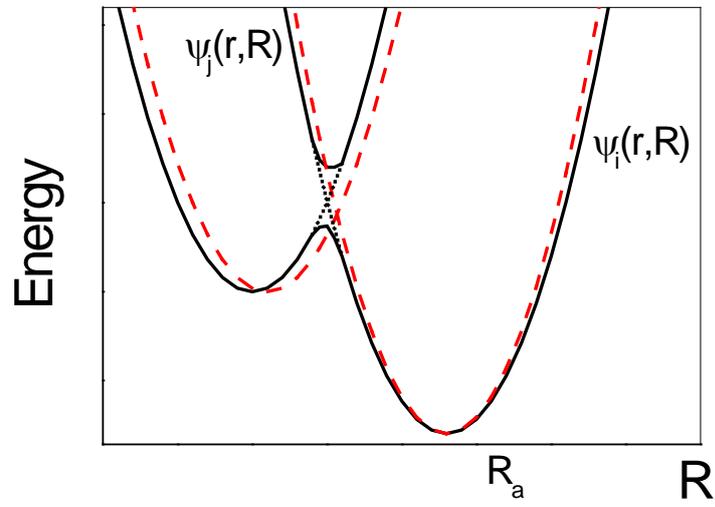

Fig 1. The coordinate diagram for static coupling theory and adabatic approximation. The dashed lines indicate $\psi_i(r, R_a)$ and $\psi_j(r, R_a)$. The dotted line connected branches are $\psi'_i(r, R)$ and $\psi'_j(r, R)$ described in the text.

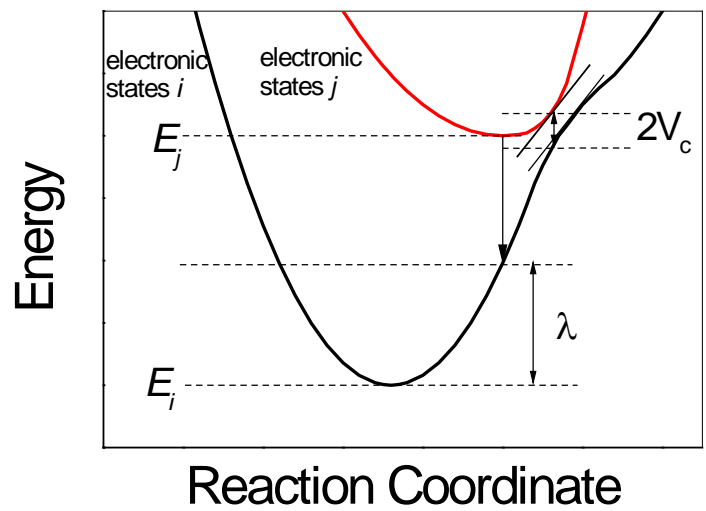

Fig 2. Marcus theory energy diagram.

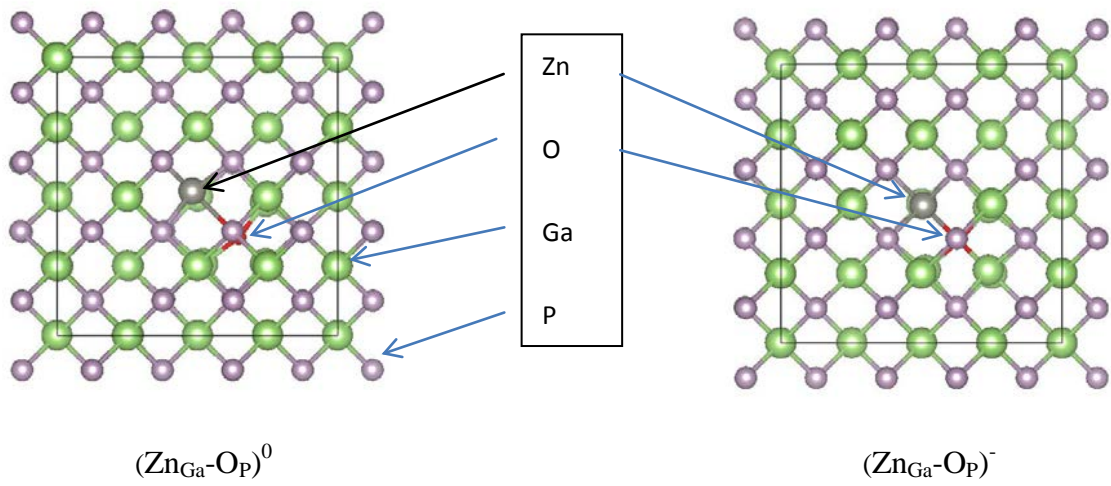

$(Zn_{Ga}-O_P)^0$          $(Zn_{Ga}-O_P)^-$

FIG. 3. The atom structures of $Zn_{Ga}-O_P$ in GaP at different charge states.

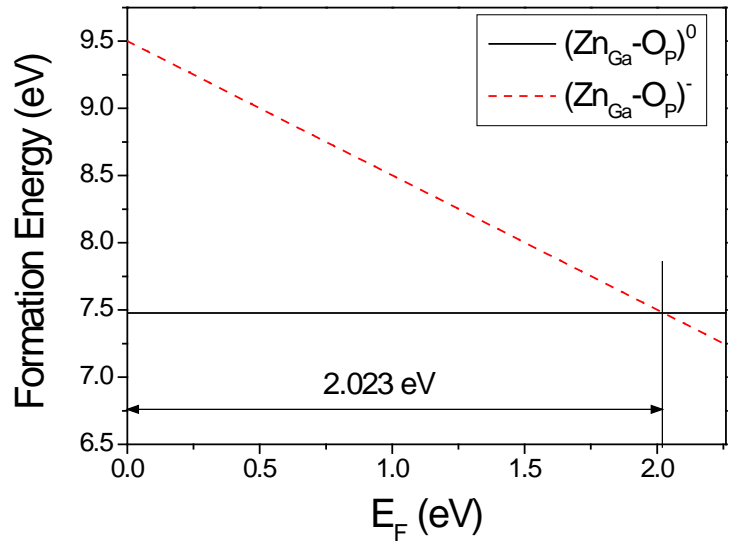

FIG. 4. The formation energy of $Zn_{Ga}$-$O_P$ in GaP at different charge states calculated under HSE. Note, the material we studied is p-type doped, thus at equilibrium it has a neutral defect. During the nonradiative recombination process, an electron first falls from the conduction band to the defect state, make it a "-" charged defect. This process is the rate determining process.

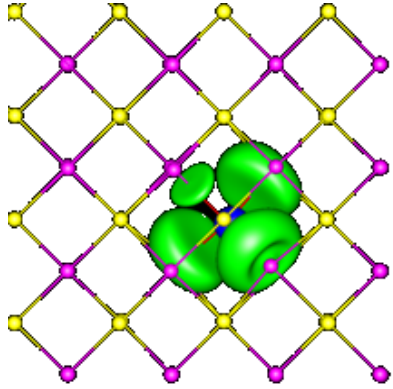

FIG. 5. The impurity wavefunction of $Zn_{Ga}$-$O_P$ center in the 64-atom supercell calculated using the HSE DFT functional.

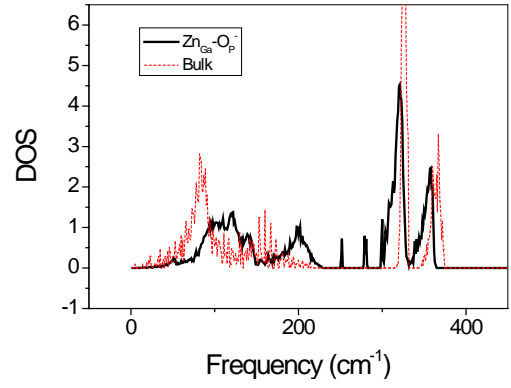

FIG. 6. The phonon spectrum, the black solid line is calculated by the CDM with $R_c$ =6.0 a.u., the red dashed line is the GaP bulk.

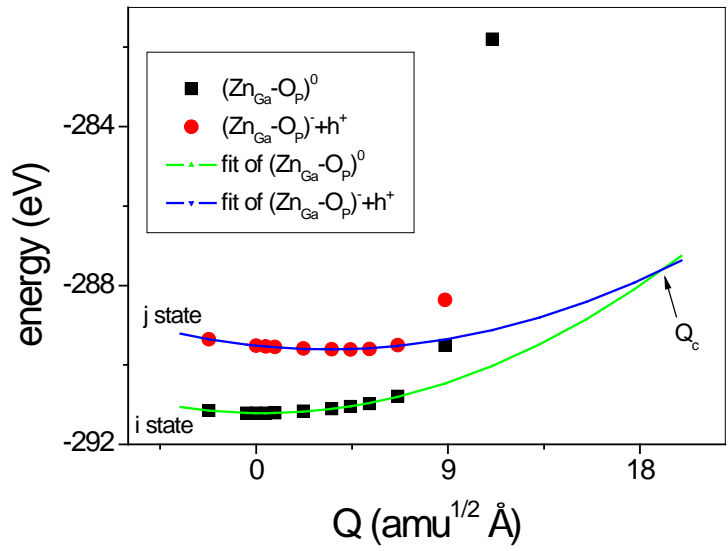

FIG. 7. Calculated 1D cc diagram for $Zn_{Ga}$-$O_P$ center in GaP. Symbols: calculated values; solid line: parabolic fit.

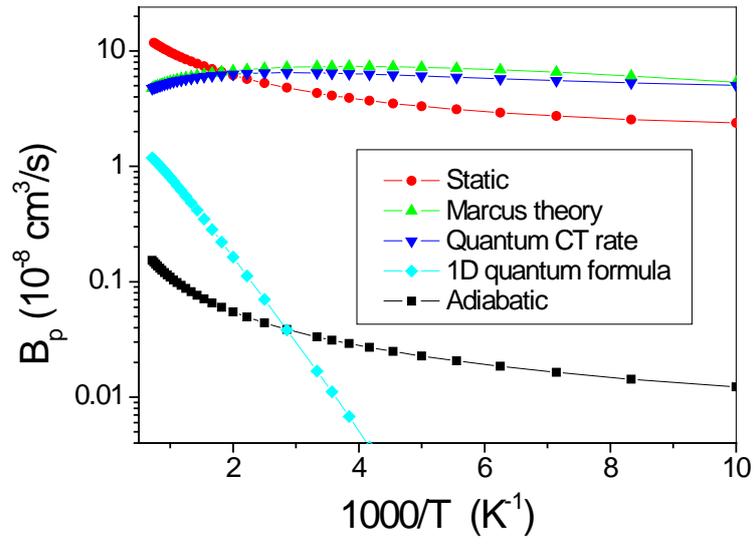

FIG. 8. The functions of the capture coefficients with 1000/T for GaP:$Zn_{Ga}$-$O_P$.

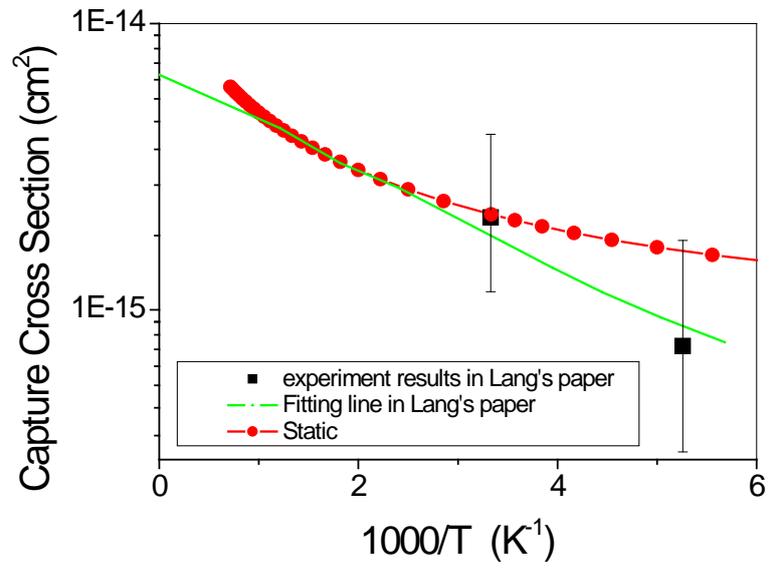

FIG. 9. The functions of the capture coefficients with 1000/T for GaP:$Zn_{Ga}$-$O_P$

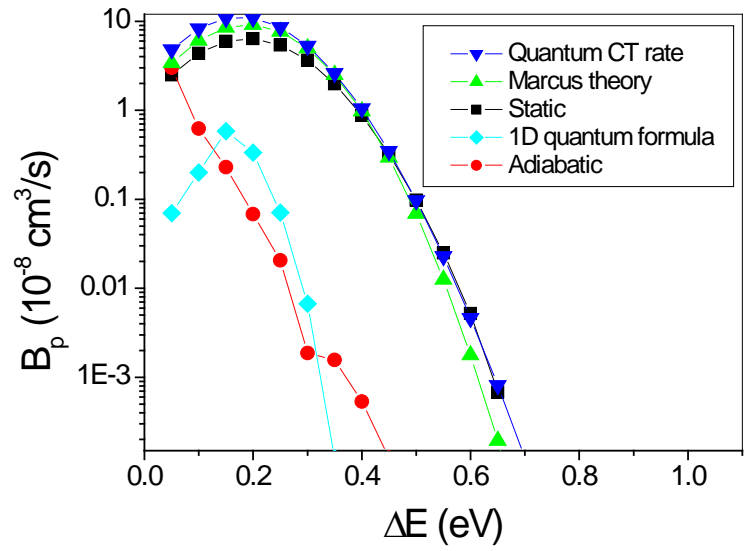

FIG. 10. The functions of the capture coefficients with $\Delta E$ for $Zn_{Ga}$-$O_P$ center in GaP.

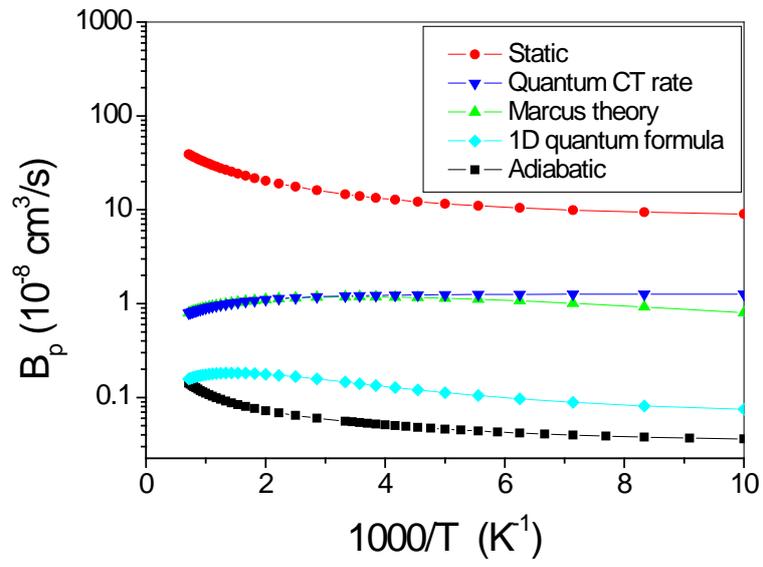

FIG. 11. The functions of the capture coefficients with 1000/T for GaN: $Zn_{Ga}$-$V_N$.

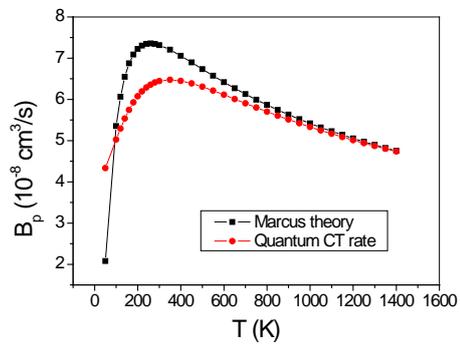 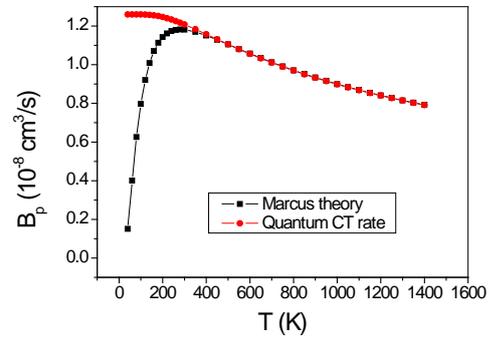

GaP: $Zn_{Ga}$-$O_P$                      GaN:$Zn_{Ga}$-$V_N$

FIG. 12. The comparing between Marcus theory and Quantum CT rate.